# Scalar Nanosecond Pulse Generation in a Nanotube Mode-Locked Environmentally Stable Fiber Laser

R. I. Woodward,[a] E. J. R. Kelleher,[a] D. Popa,[b] T. Hasan,[b] F. Bonaccorso,[b,c] A. C. Ferrari,[b] S. V. Popov,[a] and J. R. Taylor.[a]

[a] Femtosecond Optics Group, Department of Physics, Imperial College London, London SW7 2BW, U.K.

[b] Cambridge Graphene Centre, University of Cambridge, Cambridge CB3 0FA, U.K.

[c] Istituto Italiano di Tecnologia, Genova 16163, Italy).

Corresponding author: jr.taylor@imperial.ac.uk

## Abstract

We report an environmentally stable nanotube mode-locked fibre laser producing linearly-polarized, nanosecond pulses. A simple all-polarization-maintaining fibre ring cavity is used, including 300 m of highly nonlinear fibre to elongate the cavity and increase intracavity dispersion and nonlinearity. The laser generates scalar pulses with a duration of 1.23 ns at a centre wavelength of 1042 nm, with 1.3-nm bandwidth and at 641-kHz repetition rate. Despite the long cavity, the output characteristics show no significant variation when the cavity is perturbed, and the degree of polarization remains at 97%.

Laser mode locking, fibre lasers, optical polarization, optical pulse shaping.

## Introduction

Passively mode-locked fibre lasers have become established tools for generating picosecond and femtosecond pulses due to their compact design, alignment-free waveguide format and low cost [1]. These advantages over their solid-state counterparts have driven interest in extending the mode locked fibre laser format to produce highly-stable nanosecond duration pulses at sub-megahertz repetition rates [2]–[6]. Such pulse sources could simplify amplification schemes [2] and enable lower threshold-power super-continuum generation [7].

Mode-locked pulse characteristics are determined by the laser cavity dispersion map [8], [9]. In particular, all-normal dispersion cavities have been reported to produce chirped pulses with energies up to hundreds of nJ [10]–[12]. These pulses have been reported as examples of dissipative solitons [8], which are encountered in a wide range of physical systems [9]. In mode-locked lasers, the dissipative soliton formation mechanism requires a balance between gain and intensity-dependent loss, in addition to interplay between dispersion and nonlinearity [9]. Spectral filtering is also required to stabilize the pulse, often in the form of an intracavity filter [8] or arising from the limited gain bandwidth of the system [6]. More recently, this concept has been extended by elongating cavities up to kilometre scales, giving very large intracavity dispersion to produce nanosecond duration pulses [3]–[5]. Chirped fibre Bragg gratings have also been used to provide a large lumped dispersion to generate nanosecond pulses at megahertz repetition frequencies [13], [14]. Whilst it should be noted that lasers of this nature can also operate in a noisy regime producing periodic bursts of incoherent noise-like pulses [3], with appropriate cavity design and adjustment, coherent pulses can be produced with a giant linear chirp [5]. Such pulses could be directly amplified in a chirped pulse amplification scheme to avoid pulse distortion during power scaling, removing the need for a stretcher stage [1] and the linear chirp suggests compression is possible [5]. The low-repetition rate of long-cavity lasers enables higher pulse energies to be achieved, which could reduce the pump power required for supercontinuum generation, with applications in fluorescence imaging and broadband spectroscopy [7], [15].

The sensitivity of optical fibres to environmental stresses can, however, limit practical application of long-cavity mode-locked fibre lasers. Such disturbances can alter the fibre birefringence, which can unpredictably degrade laser performance [16]. Additionally, nanosecond modelocked pulse sources reported to date produce vector pulses [3]–[5], [13], [14], i.e. pulses consisting of two coupled propagation modes along orthogonal polarization axes [16]. These modes can exchange energy as the birefringence changes, causing the polarization state to evolve unpredictably over time [16]. When both group and phase velocities are equal along orthogonal polarization axes, i.e. polarization-locked vector solitons [17], the pulse propagates without evolution in its polarization state [17], [18]. A scalar pulse (i.e. pulse with a single polarization component [16]), with a fixed linear polarization state is preferred for many applications where polarization-dependent changes in the material are important, such as biological imaging [19] and micromachining [20].

Environmental stability of all-fibre lasers can be improved by using polarization-maintaining (PM) fibres [16]. PM fibres reduce the coupling efficiency between polarization modes, minimizing the influence of disturbance-induced birefringence changes [1], [16]. An intracavity polarization-selective element can also be added, so the optical field in one polarization axis dominates, leading to linearly-polarized light propagation. To date, all-PM cavities producing broad chirped pulses in the all normal dispersion regime have been demonstrated using both a nonlinear amplifying loop mirror [21] and a semiconductor saturable absorber mirror (SESAM) [22] to initiate modelocking, producing 68 ps [21] and 200ps [22] pulses, respectively. The degree of polarization (DOP), i.e. the polarized portion of the beam, was not measured in [21] and [22], and nanosecond scalar pulse formation in the normal dispersion regime has, until now, not been confirmed.

Carbon nanotubes (CNTs) and graphene have emerged as promising saturable absorbers (SAs) [23]–[25] with ultrafast recovery time [26], [27], the ability to support short pulses [28], [29], and with a number of favourable properties, such as broadband operation [30], and ease of fabrication [24] and integration [24] into all-fibre configurations [29], [31]. Broadband operation is an intrinsic property of graphene [32], while in CNTs this can be achieved using a distribution of tube diameters [30]. A variety of techniques have been implemented in order to integrate CNTs and graphene into lasers [33]. E.g., graphene can be integrated in various optical components [33], [34], with the possibility of controlling the modulation depth [34]. CNTs and graphene can also be embedded in polymer matrices that can easily be integrated into fibre lasers [33]. CNT composites can exhibit large modulation depths [29], [30], which are preferred for fibre lasers [1].

Here we report scalar nanosecond pulse generation in an environmentally stable mode-locked laser. We use an all-fibre all-PM ring cavity incorporating a CNT-SA composite to initiate mode-locking. We measure the output polarization properties to confirm scalar dissipative soliton operation, with a linear polarization state. Laser stability and energy fluctuations are also measured to illustrate the robustness of the pulse source.

## Experimental setup

To match the operation wavelength of Yb-doped fibre lasers (∼1060 nm), CNTs with ∼0.8 nm diameter are required [35], [36]. We use CNTs produced by Catalytic Chemical Vapour Deposition (CCVD) of $CH_4$ over $Mg_{1-x}Co_xO$ solid solution containing Mo oxide [37]. The absorption spectrum of the CNT powder is shown in Fig. 1 (black line) with a peak between 1030 and 1190 nm. The desired operating wavelength is also highlighted (red dashes).

To further investigate the tube diameter distribution, before and after composite fabrication, and monitor defects, we measure the Raman spectra of the purified powder at 457 nm (2.71 eV), 514.5 nm (2.41 eV), and 632.8 nm (1.96 eV). In the low frequency region, the Radial Breathing Modes (RBMs) are observed [38]–[40]. Their position is inversely related to the diameter, $d$ [38]–[40] by Pos(RBM) = $C_1/d+C_2$. We use $C_1$ = 214.4 cm$^{-1}$ and $C_2$ = 18.7 cm$^{-1}$. A variety of $C_1$ and $C_2$ have been proposed in literature, see [39], [40]. Here we use $C_1$ = 214.4 cm$^{-1}$ and $C_2$ = 18.7 cm$^{-1}$ [41], derived in [38] by plotting the resonance energy as a function of Pos(RBM)$^{-1}$ without any additional assumptions. The RBM spectra of the powders [Fig. 1(b)] (black lines) show a narrow distribution, spanning the 140–377 cm$^{-1}$ range. This corresponds to ∼0.6–1.8 nm diameter. Fig. 1(c) plots the spectra in the G region of purified CNTs (black curves). A weak D band is observed [$I(D)/I(G)$ = 0.04], indicating small number of defects [42], [43]. A polymer composite is then fabricated via solution processing. Fig. 1(a) plots the absorption spectra of the PVA (grey line), the CNT-PVA composite (red line) and the pristine CNTs (black line). The absorption of the PVA is ~1 order of magnitude lower with respect to the CNT-PVA composite in the 400–1600 nm range and thus, is negligible.

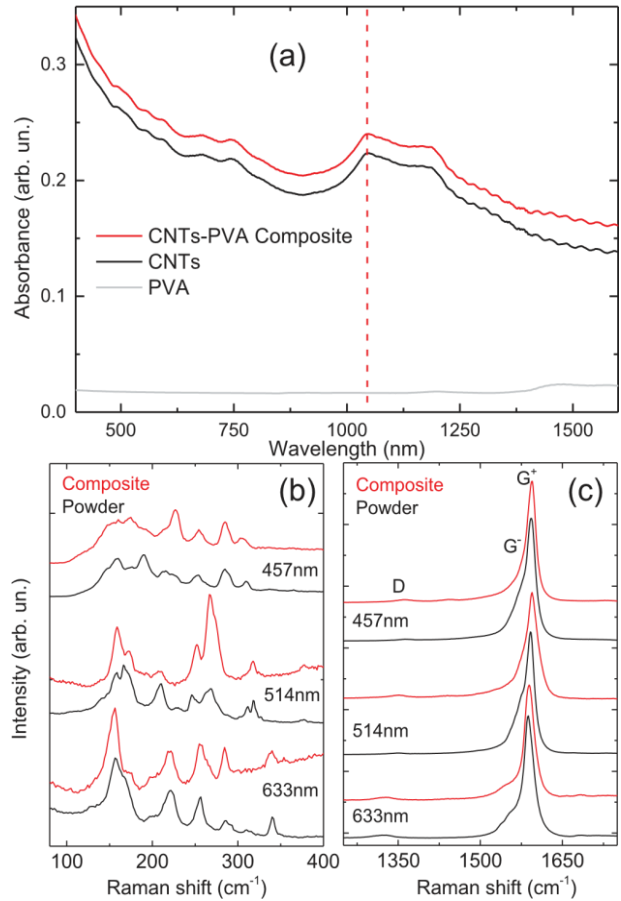

**Fig. 1**. *(a) Absorption spectrum of CNTs (black line), CNT-PVA composite (red line), and PVA (grey line). The operating wavelength is marked by the red dashed line. Raman spectra of CNT powders (black lines) and CNT-PVA composite (red lines) at different excitation wavelengths: (b) RBM region and (c) G region.*

For our laser, we use an all-fibre ring cavity design (Fig. 2). The active fibre is 0.6 m PM double-clad ytterbium doped fibre, pumped by a laser diode (LD) at 962 nm. Wavelength division multiplexers (WDMs) couple pump light into the cavity and extract un-depleted pump light. The CNT-SA is then integrated into the cavity by sandwiching the composite between two fibre connectors for angled physical contact (FC-APC). A PM isolator ensures unidirectional operation and a PM output coupler (OC) provides a 33% output. A 300 m length of PM highly nonlinear fibre (HNLF; group velocity dispersion ~24 ps$^2$/km) is used to elongate the cavity and increase the total intracavity dispersion and nonlinearity.

The remainder of the cavity consists of PM980 Panda fibre. HNLF is spliced to PM980 fibre with ~0.3 dB splice loss and >30 dB polarization extinction ratio. The total cavity length is ~312 m.

## Results and discussion

The laser exhibits self-starting mode-locking, producing a train of pulses with a 1.56 µs period, defined by the fundamental cavity frequency of 641 kHz [Fig. 3(a) inset]. The pulse energy is 0.20 nJ. Operational characteristics are unchanged when the system is intentionally perturbed by mechanically stressing the cavity fibre, demonstrating the robust modelocking performance and environmental stability of the laser.

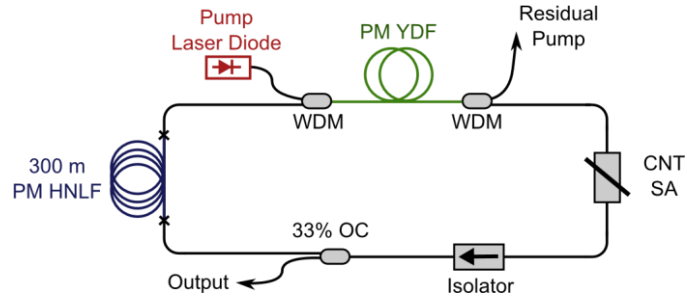

**Fig. 2.** Schematic setup of laser cavity: YDF - ytterbium-doped fibre; WDM - wavelength division multiplexer; CNT SA - nanotube-based saturable absorber; OC - output coupler; HNLF - highly nonlinear fibre.

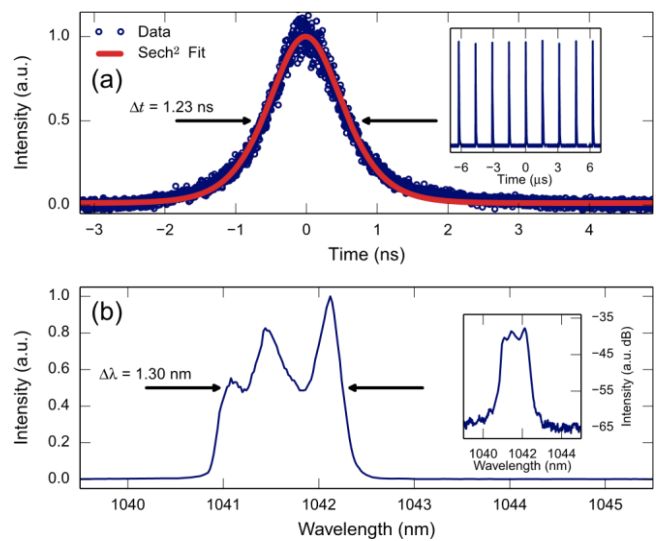

**Fig. 3.** Characteristics of laser output. (a) Streak camera trace (inset: pulse train on oscilloscope). (b) Optical spectrum on linear scale (inset: log scale).

Temporal output properties are measured using a streak camera, showing the pulse full width at half maximum (FWHM) is 1.23 ns [Fig. 3(a)]. This is, to our knowledge, the longest duration reported to date for mode-locked pulses from an all-PM fibre laser. The data are well fit by a $sech^2$ profile, as expected for pulses in long-cavity all-normal dispersion lasers [5]. Fig. 3(b) shows the corresponding output spectrum, which is centred at 1041.7 nm with a FWHM of 1.3 nm. The steep spectral edges agree with the expected spectral shape for all-normal dispersion lasers producing dissipative solitons [8]. We also observed no noise-like structures when monitoring the output with a 15 ps photodiode, confirming that the laser is not operating in a noise-burst regime. These pulse properties correspond to a time-bandwidth product (TBP) of 442, a thousand times greater than the transform-limited TBP of 0.315 [16], suggesting that the pulse is significantly chirped. Also, unlike many reported all-normal dispersion lasers [8], [10], [11], we do not require an additional spectral filter to stabilize modelocking. Instead, the limited gain bandwidth of the system, including a spectral filtering effect upon highly chirped pulses from the CNT-SA [6], could provide a stabilizing effect.

Compared to passively mode-locked nanosecond pulse sources reported to date [3]–[6], our measured spectral width of 1.3 nm is broader by more than a factor of two, despite using a much shorter cavity length. We attribute this to our use of highly nonlinear fibre to extend the cavity, rather than standard silica fibre. This increases the total cavity nonlinearity, so pulses experience greater self-phase modulation (SPM) which causes spectral broadening. Spectral broadening is then coupled to temporal broadening by the fibre group velocity dispersion. Therefore, by using HNLF we can generate nanosecond pulses with shorter cavity lengths than previously reported [3]–[6]. We suggest that adjusting cavity nonlinearity may be an alternative route to manipulating the pulse duration of all-normal dispersion lasers, preventing the need for ultralong cavities or the fabrication of chirped fibre Bragg gratings to provide large dispersion. This also highlights the need for further work to understand the complex pulse dynamics in the giant-chirped pulse regime.

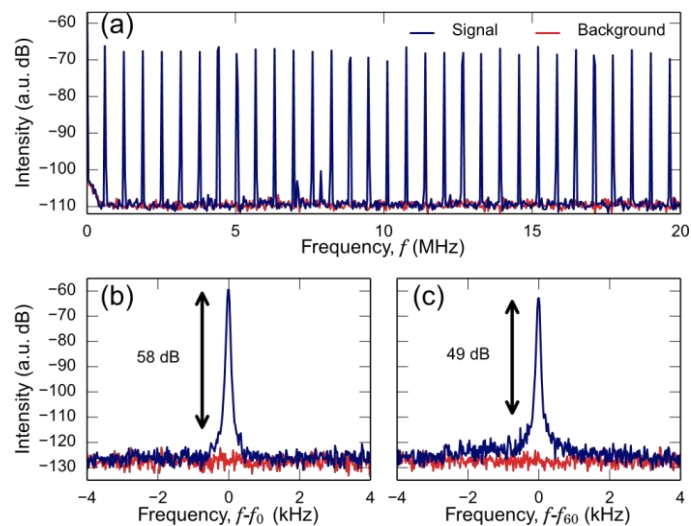

Fig. 4. RF spectrum of laser output: (a) 20 MHz span. (b) 8 kHz span about fundamental frequency, $f_0$ = 641 kHz. (c) 8 kHz span about 60th harmonic, $f_{60}$ = 38.46 MHz (30 Hz resolution bandwidth).

The polarization of our chirped pulses is monitored using a home-made polarimeter. This consists of an optical power meter placed behind a quarter wave-plate and linear polarizer (in addition to collimating optics), both of which can be rotated so that Stokes parameters can be measured. From this, we calculate a DOP of 97%, with a small ellipticity of 3 degrees. The extinction between the two polarization axes is 14.1 dB. This confirms that the output is linearly-polarized, indicating scalar pulse generation.

All previously reported nanosecond-duration mode-locked pulse sources employ a non-PM cavity [3]–[6], with a polarization controller tuned to achieve a stable mode-locking state [4], [5]. Our all-PM cavity eliminates this requirement and mode-locking is always self-starting with no need for tuning.

Stability is probed using the methods outlined in [44] by recording the RF spectrum using a 1 GHz photodetector and RF spectrum analyser. The 20 MHz span plot shows a steady train of cavity harmonics without spectral modulation [Fig. 4(a)], indicating no Q-switching instabilities [44]. We observe a high signal-to-background ratio of 58 dB ($10^{5.8}$ contrast) for the fundamental [Fig. 4(b)], which suggests good pulse-train stability. A low-level pedestal with a peak at ∼−115 dB is also detected, from which the pulse-to-pulse energy fluctuations can be estimated [44]: $\frac{\Delta E}{E} = \left(\frac{\Delta P \Delta f}{\Delta f_{rbw}}\right)^{0.5}$ where $\Delta P$ is the pedestal-peak to signal-peak power ratio, $\Delta f$ is the width of the pedestal and $\Delta f_{rbw}$ is the resolution of the spectrum analyser for this measurement. With $\Delta P = 10^{-5.8}$, $\Delta f$ = 500 Hz and $f_{rbw}$ = 30 Hz, the energy fluctuation is 0.5%. This is a low pulse energy fluctuation, especially compared to other nanosecond pulse sources, such as the 7% fluctuation reported in [14]. The 60th harmonic can also be seen to exhibit a high 49 dB peak-to-background ratio [Fig. 4(c)], further confirming the mode-locking stability.

## Conclusions

We reported the generation of scalar nanosecond-duration pulses from an environmentally stable mode-locked laser. The use of PM fibre throughout the cavity results in robust, selfstarting mode-locking, resistant to external disturbances. This paves the way to the development of mode-locked lasers generating scalar, linearly-polarized pulses with nanosecond durations and hundreds of kilohertz repetition rates for applications where reliability and stability are key requirements.

## Acknowledgements
This work was supported in part by the Royal Society Wolfson Research Merit Awards, in part by the European Research Council through the NANOPOTS, Hetero2D, European Union Graphene Flagship under Contract 604391, in part by the Engineering and Physical Sciences Research Council under Grant EP/K01711X/1, Grant EP/K017144/1, and Grant EP/L016087/1, in part by Emmanuel College, Cambridge, and in part by a Newton International Fellowship.